\documentstyle[12pt,psfig,epsfig]{article}
\newcommand{\beq}{\begin{equation}}
\newcommand{\eeq}{\end{equation}}
\newcommand{\beqarray}{\begin{eqnarray}}
\newcommand{\eeqarray}{\end{eqnarray}}
\def\lsim{\raise0.3ex\hbox{$\;<$\kern-0.75em\raise-1.1ex\hbox{$\sim\;$}}}
\def\gsim{\raise0.3ex\hbox{$\;>$\kern-0.75em\raise-1.1ex\hbox{$\sim\;$}}}
\def\para{\vspace{0.3cm}\noindent}

\def\km{\,{\rm km}}
\def\m{\,{\rm m}}
  
\def\yr{\,{\rm yr}}

\def\mpc{\,{\rm Mpc}}

\def\erg{\,{\rm erg}} 
\def\ergs{\,{\rm ergs}} 
\def\ev{\,{\rm eV}}
\def\kev{\,{\rm keV}}
\def\mev{\,{\rm MeV}}

\def\tev{\,{\rm TeV}}

\def\sec{\,{\rm sec}}

\def\Egammamax{E_{\gamma {\rm max}}}
\def\Egammamin{E_{\gamma {\rm min}}}
\def\Egrb{E_{\rm GRB}}
\def\Egamma{E_{\gamma}}
\def\Egammaobs{E_{\gamma 0}}
\def\Egammamaxobs{E_{\gamma {\rm max} 0}}
\def\Egammath{E_{\gamma {\rm th}}}
\begin{document}
\begin{center}
{\large \bf Detecting TeV Gamma Rays from Gamma Ray Bursts\\
 by Ground Based Muon Detectors}

\medskip

{Nayantara Gupta \footnote{tpng@mahendra.iacs.res.in}}\\  
{\it Department of Theoretical Physics,\\
 Indian Association for the Cultivation of Science,\\
Jadavpur, Calcutta 700 032, INDIA.}  

\bigskip

{Pijushpani Bhattacharjee\footnote{pijush@iiap.ernet.in}}\\
{\it Indian Institute of Astrophysics, Bangalore 560 034,
INDIA.}
\end{center}

\begin{abstract}
The possibility of detecting Gamma-Ray Bursts (GRBs) in TeV energy
range using large area muon detectors like AMANDA and Lake Baikal
detector is examined. These detectors can detect TeV energy photons by
detecting the secondary muons created by the TeV photons in the Earth's
atmosphere. We calculate the expected number of muons and the signal to
square root of noise ratios in these
detectors due to TeV gamma-rays from individual GRBs for various
assumption on their
luminosity, distance from the observer (redshift), gamma-ray integral
spectral index, maximum energy cutoff of the photon spectrum, 
and duration, including the effect of the absorption of TeV 
photons in the intergalactic infrared radiation background. We also
calculate the expected rate of
detectable TeV-GRB events in these detectors using a recent determination
of 
the luminosity and redshift distributions of the GRBs in the Universe. 
For reasonable ranges of values of various parameters, we find about 
1 event in 20 years in AMANDA (and a similar number for BAIKAL), while the
event rate can be significantly larger (by factors of 10 or more
depending on the area of the detector) in the proposed next generation
detectors such as ICECUBE. Although, for the specific forms of the
luminosity- and redshift distributions assumed, the average rate of
expected detection is low, occasional nearby ($z\lsim0.1$), high luminosity
($\gsim10^{54}\erg/\sec$), long duration ($\gsim$ 10's of seconds), and
sufficiently hard spectrum (gamma-ray integral spectral index $<$ 1) 
GRBs can, however, be detected even by AMANDA. Detection (or even
non-detection) of TeV photons from GRBs in coincidence with
satellite-borne detectors (which detect mainly sub-MeV photons) will be
able to provide important new insights into the characteristics of GRBs 
and their emission mechanisms and, in addition, provide limits on the
strength and spectrum of the intergalactic infrared background which
affects the propagation of TeV photons from cosmological sources. 
\end{abstract} 
\newpage
\section{Introduction}
Gamma Ray Bursts (GRBs) are among the most powerful astrophysical
phenomena in the Universe; see e.g., Ref.~\cite{grb_review} for a recent
review.  
Because of the limited sizes of the satellite-borne GRB detectors (BATSE,
BeppoSAX, for example), GRBs have been detected mostly in the sub-MeV
energy
region where the photon number flux is sufficiently large for typical
power-law photon spectra of GRBs. However, GRB emission extending to
$\sim$10 GeV is known as in the case of the famous long-duration burst GRB
940217~\cite{hurley} detected by the EGRET instrument on board the Compton
Gamma Ray Observatory (CGRO), suggesting the possibility that 
GRBs may also emit much higher energy photons, perhaps even extending to
TeV energies as in some highly energetic Active
Galactic Nuclei (AGN) of the ``blazar'' class. For
power-law spectra falling with energy, the photon number flux at TeV
energies may be too low for TeV photons from GRBs to be detected by the
satellite-borne detectors. However, ground-based detectors can in
principle detect TeV photons from GRBs by detecting the secondary
particles (electrons, muons) of the atmospheric showers generated by
these photons in  the Earth's atmosphere. A variety of techniques (see,
e.g., ~\cite{ong} for a review) allow determination of the energy
and direction of the shower-initiating primary photon with good accuracy. 

\para 
Indeed, three major ground-based gamma ray
detectors, the Tibet air shower array~\cite{tibet}, the HEGRA-AIROBICC
Cherenkov array~\cite{hegra} and the Milagro water-Cherenkov
detector~\cite{milagro} have independently claimed evidence for
possible TeV $\gamma$-ray emission from sources in directional and
temporal coincidence with some GRBs detected by BATSE. The estimated
energy in TeV photons in each case has been found to be about 1 to 2
orders of magnitude larger than the corresponding sub-MeV energies
measured by BATSE. Since TeV photons are efficiently absorbed in the
intergalactic infrared (IR) background due to pair
production~\cite{stecker-dejager}, only relatively close by (i.e., low
redshift) GRBs, for which the absorption due to IR background is
insignificant, can be observed at TeV energies, which would explain the 
fact that only a few, not all, of the
BATSE-detected GRBs in the fields of view of the individual ground
detectors have been claimed to be detected at TeV energies.      

\para 
While more GRBs would have to be detected in TeV energies before any firm
conclusions can be drawn, the above discussions already point to the 
possibility that not only may the spectra of some, if not
all, GRBs extend to TeV energies, but that the energy emitted in TeV
photons may even constitute the dominant part of the
total energy emitted by individual GRBs. 
This would happen, for example, if the photon spectrum, extending to TeV
region, is hard with a differential power-law index $\gamma<2$. 
In such situations, while the total {\it number} of photons in the entire
spectrum is dominated by those at the lower cutoff energy of the
spectrum, the total {\it energy} is dominated by the upper cutoff energy. 
In other words, it is possible that the total burst energies
of individual GRBs calculated from the measured fluences in the
keV -- MeV energies by the satellite detectors may only be a small
fraction of the actual total energies of the individual bursts, most of    
which could be carried away by the relatively few TeV photons rather than
by the abundant sub-MeV photons. 

\para 
Indeed, photon differential spectrum with power-law index $\gamma<2$ would
be expected in most situations if the radiation arises from
synchrotron radiation of power-law distributed particles (electrons or
protons): Recall that the power-law index of the synchrotron photon
differential spectrum is $(p+1)/2$ where $p$ is the power-law index of
differential number spectrum of the particles (electrons or protons). 
Thus, one can expect $\gamma<2$ as long as $p<3$, which is typically the
case for particles accelerated in shocks. 

\para 
On the theoretical side, it has been suggested~\cite{totani1,totani2} that
if protons are accelerated to energies above $\sim10^{20}\ev$ in GRBs as
envisaged in the scenario~\cite{grb-uhecr} in which GRBs are the sources
of the ultrahigh energy cosmic rays extending to energies above
$10^{20}\ev$, then the synchrotron radiation of those ultrahigh energy
protons in the magnetic field within the GRB would produce TeV photons.
Thus, while the synchrotron radiation of electrons produce the keV--MeV
flux, the TeV flux would be produced by synchrotron radiation of the
protons with energy above $10^{20}\ev$. Whether or not a GRB would be
bright in TeV would, in this scenario, be determined by the efficiency of
energy transfer from protons to electrons within the GRB~\cite{totani2}. 
It has also been suggested~\cite{vazquez,totani3} that while the TeV
photons emitted by GRBs at large redshifts would be absorbed through
$e^+e^-$ pair production on the intergalactic infrared background, the
resulting electromagnetic cascades initiated by the produced pairs could
produce the observed extragalactic diffuse gamma ray background in the GeV
energy region. It is thus clear that confirmation of TeV gamma ray
emission from GRBs would have major implications not only for the
physics and astrophysics of GRBs, but also for the strength and redshift
evolution of the intergalactic infra-red background. 

\para 
High energy primary gamma rays above a few hundred GeV create
``air-showers'' of secondary particles by interacting with Earth's  
atmosphere. Among the secondary particles are muons which can be detected,
and the energy and direction of the shower initiating parent photon
reconstructed (to a directional accuracy of typically a degree), by using
the relatively shallow underground muon
detectors such as AMANDA~\cite{amanda}, Lake Baikal detector~\cite{baikal}
and the proposed ICECUBE~\cite{icecube}, and surface muon detector such as
MILAGRO~\cite{milagro}. The first three are facilities 
primarily for detecting high energy neutrinos from cosmic sources ---
the neutrinos are detected by detecting the muons produced by the
neutrinos in ice (AMANDA and ICECUBE) or water (Lake Baikal) --- while
MILAGRO is a water-Cherenkov detector of ``all'' particles (including
muons) primarily for detecting gamma rays above a few hundred GeV from
astronomical sources. 

\para 
The advantages of using these muon detectors over the conventional
air-Cherenkov telescopes for doing TeV gamma ray astronomy have been
expounded in Refs.~\cite{hsy,ah}. Compared to air-Cherenkov telescopes,
the muon detectors cover much larger fraction of the sky with large duty
cycles. For example, the AMANDA detector at the South Pole covers more
than a quarter of the sky with essentially 100\% efficiency. 
Being at
relatively shallow depths --- of order 1 km --- the ``neutrino'' detectors
are sensitive to muons with energies of a few hundred GeV and thereby to
parent photons of energy of order a TeV and above. MILAGRO detector's
muon detection threshold is smaller --- about 1.5 GeV --- because of 
its location on the surface, which makes it sensitive to even lower energy
gamma rays.

\para 
As pointed out in Ref.~\cite{ah}, these muon detectors are particularly
suited for detecting TeV gamma rays from transient sources like GRBs ---
the otherwise large background of down-going cosmic-ray induced
atmospheric muons can be significantly reduced because, with the
information on the time and duration of a burst provided by satellite
observation, the background is integrated only over the relatively short   
duration of a typical GRB which is of order few seconds.

\para 
In this paper we calculate, following Ref.~\cite{ah}, the expected number
of muons and the signal to square root of noise ratios in AMANDA and Lake
Baikal detectors due to TeV gamma-rays from individual GRBs for various
assumption on their 
luminosity, distance from the observer (redshift), gamma-ray integral
spectral index, maximum energy cutoff of the photon spectrum, 
and duration, including the effect of the absorption of TeV 
photons in the intergalactic infrared radiation background. We
carefully take into account all relevant cosmological effects (in a
cosmological constant dominated present Universe, see below for details).
Our results for number of muons for individual GRBs differ significantly
(especially in the case when absorption in the intergalactic IR background
is taken into account) from those obtained in Ref.~\cite{ah} for the same
set of GRB parameters. We also calculate the expected rate of 
detectable TeV-GRB events in AMANDA-like detectors using a 
recent determination of the luminosity and redshift distributions of the
GRBs in the Universe. 

\para 
This paper is organized as follows: 
In Sec.~2 we discuss the parametrization of the TeV emission from
the GRBs and the formalism we have
used to calculate the number of secondary muons produced in the atmosphere
from these TeV gamma rays of the GRBs. In Sec.~3 the results of our
calculation of the number of muons expected and the signal to square root
of noise ratios due to TeV photons from individual GRBs as functions of
various GRB parameters are discussed. The expected rate of detection of
TeV GRBs is calculated in Sec.~4 by using a recently determined luminosity
function and redshift distribution of the GRBs in the Universe, with
summary and conclusions presented in Sec.~5. 

\section{Parametrizations of the Photon Spectrum of GRBs and the 
Secondary Muons in Gamma-Ray Induced Showers in Earth's Atmosphere}

\subsection{Photon Spectrum of GRBs}
We shall assume the emission spectrum of the GRB to be constant
during the emission time $\delta{t_{e}}$, and represented by
$dN_\gamma/d\Egamma = K\Egamma^{-\alpha-1}$, giving the total
number of photons emitted during the entire burst per unit energy at
energy $\Egamma$ between a minimum and a maximum energy, $\Egammamin$ and
$\Egammamax$, respectively. Here $\alpha>0$ is the integral
spectral index and $K$ is a normalization constant, which are related
to the total energy emitted from the burst, $\Egrb$, by   
\beq  
\Egrb = K\int_{\Egammamin}^{\Egammamax}
\Egamma^{-\alpha}d\Egamma\,,\label{e_grb}
\eeq
from which we get
\beq
K = \Egrb\times\left\{ \begin{array}{ll}
\left(-\alpha +1\right)\left(\Egammamax^{-\alpha + 1} -
\Egammamin^{-\alpha + 1}\right)^{-1} & \mbox{for $\alpha\neq 1$}\\
\left[\ln\left(\Egammamax/\Egammamin\right)\right]^{-1} &
\mbox{for $\alpha=1\,.$}
\end{array}
\right. \label{K}
\eeq
Following Ref.~\cite{ah}, we shall use $\Egammamin=1\mev\, $ since
negligible amount of energy is emitted below that energy, and assume that 
the spectrum continues to TeV energies with the same integral spectral
index $\alpha$. Further, as discussed in the Introduction, we shall
assume the photon spectrum to be sufficiently hard with differential
spectral index $\gamma<2$, i.e., integral spectral index $\alpha<1$.  

\para 
The luminosity, $L$, and the total energy emitted, $\Egrb$, are related as
\beq
\Egrb=L\times \delta t_e\,.\label{L}
\eeq
For a GRB source located at a cosmological distance corresponding to a
redshift $z$, the observed burst duration is
$\delta{t}=(1+z)\delta{t_e}$. For our purpose, a GRB is completely
specified by the parameters $L,\, z,\, \delta{t_e},\, \alpha\, $ and
$\Egammamax$. 

\para 
The observed spectrum is determined
by two factors: (a) the photons energies are redshifted by a factor
$(1+z)$ due to the expansion of the universe, and (b) GeV -- TeV photons 
are absorbed in the intergalactic space due to $e^+e^-$ pair production 
on the intergalactic starlight and infrared background photons. 
The probability for a photon emitted with energy $\Egamma$ by a source at
redshift $z$ to reach the earth is 
$e^{-\tau(\Egamma,z)}$, where $\tau$ is the optical depth for absorption
due to the pair production process. We have used the optical depths given
in \cite{stecker-dejager}, where a parametrization has been given for
$\tau(\Egamma,z)$ for small $z$ which, as we shall see, is appropriate for
our purpose. 

\para 
The observed photon energy $\Egammaobs$ and its energy at the source,
$\Egamma$, are related as $\Egammaobs=\Egamma/(1+z)$. Thus, at 
the top of the atmosphere, the observed cutoff energy of the source is
$\Egammamaxobs=\Egammamax/(1+z)$. Assuming isotropic
emission from the source, the total number
of photons from a single GRB at redshift $z$ per unit energy at energy
$\Egammaobs$ that strike per unit area of the Earth at the top of the
atmosphere is 
\beq
\frac{d\phi_0}{d\Egammaobs}=\frac{K}{4\pi r^2(z)}
\Egamma^{-\alpha-1}(1+z)e^{-\tau(\Egamma,z)}\,,\label{dphide}
\eeq 
where $r(z)$ is the comoving radial coordinate distance of the source. 
For a spatially flat Universe (which we shall assume to be the case) with
$\Omega_\Lambda + \Omega_m = 1$, where 
$\Omega_\Lambda$ and $\Omega_m$ are, respectively, the
contribution of the cosmological constant and matter to the energy density
of the Universe in units of the critical energy density, $3H_0^2/8\pi G$, 
the radial coordinate distance $r(z)$ is given by~\cite{mtw}  
\beq
r(z)=\int^{z}_{0}\left(\frac{c}{H_0}\right)\frac{dz^\prime}
{\sqrt{\Omega_{\Lambda}+ \Omega_m(1+z^\prime)^3}}\,.\label{r_z}
\eeq
Here $c$ is the speed of light, $H_0$ is the Hubble constant in the
present epoch. In our calculations, we shall use  
$\Omega_\Lambda=0.7$, $\Omega_m=0.3$, and $H_{0}=65\, \km\, \sec^{-1}\,
\mpc^{-1}$.

\subsection{Number of Muons in a Photon-Induced Air Shower} 
Photons of sufficiently high energy striking the earth's atmosphere 
interact with the air nuclei to produce 
pions; these photo-pions are the major source of muon production
by photons in atmospheric air-showers, among other possible physical
processes like direct pair production of muons by photons or
photo-produced charm decays. The number of muons 
with energy above $E_{\mu}$ in an air-shower produced by a single photon
of energy $\Egammaobs$ striking the atmosphere can be
parametrized~\cite{muon,ah}, for $E_\mu$ in the range 0.1 TeV to 1 TeV, by
the formula 
\beq
N_{\mu}(\Egammaobs,\,
>E_\mu)\simeq\frac{2.14\times10^{-5}}{\cos\theta}\frac{1}
{(E_\mu/\cos\theta)}\frac{\Egammaobs}{(E_\mu/\cos\theta)}\,,
\label{muon_single_photon} 
\eeq
where $\theta$ is the zenith angle of the photon source, and all energies
are in TeV units. 

\para 
The above parametrization is valid for $\Egammaobs/E_\mu\geq10$ \cite{ah},
which is adequate for the purpose of calculating the muon rates in
the AMANDA and Baikal detectors both of which have muon thresholds of the 
order of a few hundred GeV~\cite{ah}.  

\para 
The total number of muons with energy in excess of
$E_{\mu}$ in a detector of effective area $A$ due to a single GRB at
redshift $z$ can then be written as 
\beq
N_\mu(>E_\mu)=A\int_{E_{\gamma {\rm th}}}^{\Egammamaxobs(z)} 
d\Egammaobs\frac{d\phi_0}{d\Egammaobs}(z)N_\mu(\Egammaobs,\,
>E_\mu)\,,\label{muon_all1}
\eeq
where $E_{\gamma {\rm th}}\simeq{10\times E_\mu/\cos\theta}$ is the
minimum photon energy needed to produce muons of energy $E_\mu$ in the
atmosphere~\cite{ah}, $\theta$ being the zenith angle of the photon
source. Using Eq.~(\ref{dphide}) in (\ref{muon_all1}) we finally have  
\beq
N_\mu(>E_\mu)= A\frac{K(1+z)^{-\alpha}} 
{4\pi r^2(z)}\int_{E_{\gamma {\rm th}}}^{\Egammamaxobs(z)}d\Egammaobs\,  
\Egammaobs^{-(\alpha+1)}e^{-\tau\left(\Egammaobs(1+z),z\right)}
N_\mu(\Egammaobs,\, >E_\mu)\,,\label{muon_all2}
\eeq
where $K$ is a function of the GRB parameters 
$L,\, z,\, \delta{t_e},\, \alpha\, $ and
$\Egammamax$, and is given by Eqns.~\ref{K} and \ref{L}. 
\section{TeV Photon Signal of GRBs in AMANDA and Lake Baikal  
Muon Detectors}
The threshold muon energy for detection of
vertical muons in the AMANDA detector~\cite{amanda} of effective area
$A\simeq10^4\m^2$ is $\sim$ 350 GeV. For the Lake Baikal
detector~\cite{baikal} the vertical muon 
threshold energy is $\sim$ 150 GeV and the effective area is 
$\sim10^3\m^2$. 
The signal in the detector, $S=N_\mu(E_\mu)$, must be compared with the
square root of noise, $\sqrt{N}$, given by  
\beq
\sqrt{N}=\sqrt{A\times\delta{t}\times
I_\mu(\theta)\times\delta\theta^2}\,, \label{noise}
\eeq
where $\delta{t}$ is the observed duration of the burst, $I_\mu(\theta)$
is the background muon intensity (i.e., number of muons per unit area,
time and solid angle) due to cosmic ray induced atmospheric
showers~\cite{amanda,ah}, and  
$\delta\theta$ is the angular resolution of the detector which is
typically a few degrees. The smallest value of $n\equiv S/\sqrt{N}$
necessary to consider an excess signal as a positive detection depends on
the search strategy of the particular detector. 
\vskip 3cm
\begin{center}
TABLE I\\
Expected number of muons in AMANDA detector for a GRB at $z=0.1$ 
\vskip 0.5cm
\begin{tabular}{|c|c|c|c|c|c|c|}
\hline
$L$ & $\Egammamax$ & $\alpha$ & No. of & $S/\sqrt{N}$ & No. of  &
$S/\sqrt{N}$ \\
(ergs/sec) & (TeV) & & muons & & muons & \\
&&& $\theta=0^{\circ}$ & $\theta=0^{\circ}$ & $\theta=60^{\circ}$ 
& $\theta=60^{\circ}$\\
\hline
$10^{52}$&10.&1.0&0.06&0.214&0.0034&0.03\\
$10^{52}$&10.&0.8&0.19&0.635&0.01&0.1\\
$10^{52}$&10.&0.5&0.38&1.276&0.026&0.23\\
\hline
$10^{52}$&5.&1.0&0.03&0.103&0.0&0.0\\
$10^{52}$&5.&0.8&0.09&0.32&0.0&0.0\\
$10^{52}$&5.&0.5&0.22&0.744&0.0&0.0\\
\hline
$10^{54}$&10.&1.0&6.404&21.404&0.348&3.16\\
$10^{54}$&10.&0.8&19&63.55&1.13&10.31\\
$10^{54}$&10.&0.5&38.18&127.68&2.6&23.7\\
\hline
$10^{54}$&5.&1.0&3.09&10.33&0.0&0.0\\
$10^{54}$&5.&0.8&9.727&32.52&0.0&0.0\\
$10^{54}$&5.&0.5&22.28&74.46&0.0&0.0\\
\hline
\end{tabular}
\end{center}
\vskip 1cm

\para 
In Table I we present for illustration the result of our calculations of
the number of
muons expected in AMANDA and the corresponding signal to
square root of noise ratios ($S/\sqrt{N}$) due to a single GRB at redshift 
$z=0.1$, for various possible values of its source luminosity ($L$),
maximum energy cutoff at
source ($\Egammamax$), and integral spectral index ($\alpha$).   
The duration of the burst is taken to be 
${\delta}t=11\sec$, somewhat intermediate between short- and long  
duration bursts. The angular  
resolution of the AMANDA detector has been taken to be
$\delta\theta=1^{\circ}$. 
The muon numbers are given for two different values of the zenith angle
$\theta$ of the source. 
The absorption of the TeV photons in the intergalactic infrared
and optical backgrounds has been taken into account in calculating the
numbers shown in Table I for which we have taken the optical depths
corresponding to the higher level of the intergalactic infrared
radiation field (IIRF) given in Ref.~\cite{stecker-dejager} in order to
have conservative estimates of the number of muons produced. Throughout
this paper we use optical depths corresponding to this higher level of
IIRF whenever we include the effect of absorption of the TeV photons in
the intergalactic space.  

\para 
Clearly, the number of muons (``signal'') and the signal-to-noise ratios
($S/\sqrt{N}$) depend
crucially on the various GRB parameters. Some of these dependencies are
illustrated in more details in Figures below. As pointed out in \cite{ah},
for a given set of other parameters, the smaller number
of muons at the larger zenith angle is a consequence of the increase of
the energy threshold with zenith angle because the muons have to traverse
a greater amount of matter to reach the detector. Indeed, for too low
value of the source gamma-ray energy cutoff $\Egammamax$, for
sufficiently large zenith angle, the threshold gamma-ray energy
$\Egammath$ required to produce muons above the threshold energy of the
detector can be larger than $\Egammamax$, thus yielding no muons. 
Also, as expected, for a
given luminosity $L$ and spectral index $\alpha\leq1$, higher source
cutoff energy $\Egammamax$ yields higher number of muons as well as higher
$S/\sqrt{N}$. However, the effect of decrement of the signal due to
smaller $\Egammamax$ can, in some cases, be over-compensated for by a
harder source spectrum (i.e., smaller value of $\alpha$) yielding a
relatively larger signal.    

\para 
Fig.~1 and Fig.~2 show the number of muons expected in AMANDA  
as a function of gamma-ray integral spectral index of the GRB
source assuming $L=3\times10^{52}\ergs/\sec$, ${\delta}t_e=1\sec$ (so that
$\Egrb=3\times10^{52}\ergs$), $\Egammamax=10\tev$, and $z=0.1$, for zenith 
angles $\theta=0^{\circ}$ (Fig.~1) and
$\theta=60^{\circ}$ (Fig.~2). The results for the BAIKAL detector are
shown in Fig.~3. Curves are shown for both without and  
with absorption of the TeV photons in the IIRF. 
In Fig.~1 and Fig.~2, the results of Ref.~\cite{ah} are also shown for
comparison. 

\para 
From Fig.~1 and Fig.~2, we see that there are significant differences
between our results and those of
Ref.~\cite{ah} for the same set of GRB parameters, with our results
for the number of muons being always consistently lower than those of
Ref.~\cite{ah}. In the case of absence of absorption, we guess the
difference between our results and those of \cite{ah} could be 
due to our using a specific cosmological model --- as already mentioned,
we use the ``standard'' cosmological model with a spatially flat Universe
with $\Omega_\Lambda=0.7\, $,
$\Omega_m=0.3\, $ and $H_0=65\km\sec^{-1}\mpc^{-1}$ --- whereas
Ref.~\cite{ah} does not explicitly mention the values of various
cosmological parameters used. 

\para 
In the case when absorption of TeV photons in the IIRF is included, 
the difference between our results and those of \cite{ah} is even more
than that in the case of no absorption. The flux of the TeV
photons, and hence the number of muons in the detector, depend quite
sensitively on the optical depth of the TeV photons, which in turn is
governed by the strength of the IIRF. A relatively higher (lower) level of
IIRF gives relatively larger (smaller) optical depth, resulting in
relatively smaller (larger) number of muons in the detector. Since the
level of IIRF is not known with certainty, we have used optical depths
corresponding to the higher level of IIRF from 
Ref.~\cite{stecker-dejager} in order to obtain conservative estimates of
the number of muons in the detector. However, Ref.~\cite{ah} does
not explicitly mention whether the optical depths corresponding to the
higher or the lower level of the IIRF given in Ref.~\cite{stecker-dejager}
was used in their actual calculation, and so it is difficult to find the
exact reasons for the differences between our results and those of
Ref.~\cite{ah}. For redshift $z=0.1$ our results differ from the
results of Ref.~\cite{ah} by almost a factor of ten in the case when
the effect of absorption of the TeV photons in the IIRF is included.  

\para 
The above discussion demonstrates the fact that the detectability of
possible TeV gamma-rays from GRBs in ground based muon detectors depends
sensitively on the various cosmological parameters and the strength and
spectrum of the IIRF in addition to the intrinsic GRB parameters.
Conversely, within the context of the ``standard cosmological model'', the
detection or non-detection of TeV photons from GRBs can significantly
constrain the strength and spectrum of the IIRF, provided the intrinsic
GRB parameters such as their total luminosity, redshift and duration are
known with reasonable accuracy from independent observations. 

\para 
In Fig.~4. the number of muons expected in AMANDA from a GRB is shown as a
function of the redshift of the GRB for $\Egammamax=10\tev$ and for two
different cases, namely, $L=3\times10^{52}\ergs/\sec$ , $\alpha=1.0$,
and $L=10^{54}\ergs/\sec$, $\alpha=0.5$. The  absorption of TeV photons 
is included with optical depth corresponding to the
higher level of IIRF from \cite{stecker-dejager}. As expected, the number
of muons from a GRB falls steeply with increasing redshift due to
absorption of TeV photons, and so detection is possible only for 
relatively nearby (low redshift) bursts.  
%At higher redshifts the discrepancy between our results and that obtained 
%in \cite{Ja} increases which is due to the difference in  the values of 
%optical depths used in the calculation.  

\para 
Fig.~5. shows the expected number of muons (for $\theta=0^{\circ}$) in
AMANDA as a function of
the total luminosity of the GRB for a fixed duration  
${\delta}t=1\sec$ of the burst. In Figs.~6--8 we show the effect 
of varying the physical parameters of a GRB on the signal 
to square root of noise ratios in the AMANDA detector. 
It is clear that high signal to square root of noise detection of TeV
photons from GRBs in AMANDA (or other ground based muon detectors of
similar area and energy threshold) will be possible mostly for  
relatively nearby ($z\lsim0.1$), long duration ($\gsim$ several tens of
seconds), high luminosity ($\gsim {\rm few}\, 
\times10^{52}\ergs/\sec$) and hard spectrum ($\alpha<1$) GRBs. 
\vskip 1cm
\begin{center}
Table II\\
\vskip 0.5cm
\begin{tabular}{|c|c|c|c|c|}\hline
BATSE&50--300 keV Luminosity&Redshift&Burst duration &1
MeV--10 TeV Luminosity \\
Number&($L_L$) from \cite{fen} (ergs/sec)&$z$&($\delta t$)
(sec)&($L_H$)
(ergs/sec) \\
\hline         
6707&$2.63\times10^{45}$&0.0085&20.6&$2.48\times10^{48}$\\
2123&$0.18\times10^{50}$&0.1&22.0& $10.06\times10^{51}$\\
2316&$0.64\times10^{49}$&0.1&29.2&$7.58\times10^{51}$\\
3055&$0.61\times10^{50}$&0.2&40.6&$2.12\times10^{54}$\\
3352&$0.62\times10^{50}$&0.2&46.3&$1.86\times10^{54}$\\
4368&$6.04\times10^{51}$&0.4&36.5&$0.88\times10^{60}$\\
\hline
\end{tabular}
\end{center}

\para 
In Table II we list some of the nearby GRBs from the BATSE
catalogue whose luminosities in the 50 to 300 keV photon energy band are
known and whose redshifts have been estimated in Ref.~\cite{fen} by using
a conjectured variability -- luminosity relationship for GRBs 
which is calibrated using seven GRBs whose redshifts are known
independently from afterglow studies. 
We have calculated the luminosities for photon energies in
the range 1 MeV to 10 TeV required for these GRBs to produce
$S/\sqrt{N}= 1.5$ at zenith angle $\theta=0^{\circ}$ in AMANDA. 
We assume $\alpha=1.0$ for these GRBs. The luminosities so calculated, 
denoted by $L_H$, are displayed in the last column of Table II.  
The intergalactic absorption of TeV photons corresponding to the higher IIRF
level from \cite{stecker-dejager} has been taken into account in this 
calculation.
 
\para 
From Table II we see that the high energy (1 MeV to 10 TeV) luminosities
($L_H$) required to detect TeV photons from these GRBs in AMANDA are much
higher than their luminosities $L_L$ in the lower energy (50 keV to 300
keV) band. No TeV photons from these GRBs have been reported to be
detected by AMANDA. The most likely explanation is that these GRBs did
not emit sufficient number of TeV photons that would create
sufficient number of muons above the threshold of AMANDA and/or the
intergalatic absorption of TeV photons is even stronger than what we
assumed. 
\section{Average Rate of Detectable TeV GRBs in Muon Detectors}
In the previous sections we have studied the detectability of TeV photons
from individual GRBs in muon detectors by calculating the number of muons
produced and the signal to noise ratios as functions of various intrinsic
parameters of the bursts such as their luminosities, integral spectral
index, redshift, and burst duration.  

\para 
In this section we estimate the expected rate of positive detection of 
TeV photons from all GRBs in the Universe in muon detectors such as
AMANDA. This requires the knowledge of the forms of the luminosity
function (LF) and the redshift distribution (RD) of the GRBs in the
Universe. 
Unfortunately, direct measurements of redshifts exist for only a few
GRBs (from optical afterglow measurements), and so the RD 
and the LF of the GRB population are not known with
certainty. Nevertheless, recently there have been attempts to indirectly
derive the redshifts (and hence luminosity) distributions of GRBs from the 
BATSE data exploiting various features of the light curves of individual
GRBs in the BATSE data. 
\subsection{Luminosity Function and Redshift Distribution of GRBs}
Recently, authors of Ref.~\cite{fen} have found that the
luminosities of seven GRBs in the BATSE catalogue with known redshifts
(measured from their optical afterglows) are correlated (as a power law) 
with the variability of the individual bursts, 
the variability being defined as the normalized 
variance of the observed 50--300 keV light curve about a fitted smooth
light curve. Based on this observation, the authors of Ref.~\cite{fen}
have conjectured a variability ($V$)--luminosity ($L$)
relationship which is hypothesized to be true for all GRBs in the BATSE
catalogue. Use of this conjectured $V$--$L$ relationship, which is
calibrated using the seven GRBs with known redshifts, then allowed
estimation of the redshifts of about 220 GRBs in the BATSE catalogue,
which in turn allowed construction of a LF and a RD for GRBs in general.    

\para 
Independently, authors of
Ref.~\cite{norris} have found a correlation between luminosity and the
so-called ``lag'' ($\tau_{\rm lag}$), which is the time delay between the
peaks in the light curves of individual GRBs. It was then pointed out in
Ref.~\cite{deng} that the above two relationships, namely the $V$--$L$ and
the $\tau_{\rm lag}$--$L$ relationships, if true in general, would
together imply a $V$--$\tau_{\rm lag}$ relationship which could be tested
directly on the measured data for BATSE observed GRBs without reference to 
their redshifts. Indeed, Ref.~\cite{deng} claims strong concordance of the
above mentioned $V$--$\tau_{\rm lag}$ relationship with the
observed BATSE data for 112 GRBs, based on which a LF has been 
derived for GRBs in general, which is a broken power-law with a break at
$L\approx 2\times10^{52}\erg/\sec$, with LF going as $L^{-1.7\pm0.1}$
below the break and $L^{-2.8\pm0.2}$ above the break. The RD 
giving the number density of GRBs as a function of redshift is also
derived~\cite{deng}, which goes as $(1+z)^{2.5\pm0.3}$. 

\para 
We must remember that the LF mentioned above refers to the luminosity
in the BATSE band (50--300 keV), which we have denoted above by $L_L$. We
are, however, interested in the LF expressed as a function of the
luminosity in the 1 MeV --- 10 TeV interval, which we have denoted above
by $L_H$. It is, however, easy to see that, for a single power-law photon
spectrum across all energies, which we assume for simplicity in this
paper, the LF expressed as a function of $L_H$ has the same functional
form as that in terms of $L_L$, except that the break of the LF
occurs at a different value of luminosity which depends on the integral
spectral index $\alpha$. For a given GRB with a definite
$L_L$ and $L_H$, one can see that the two are related as 
\beq
\frac{L_H}{L_L}= \frac{(10\tev)^{-\alpha +1} - (1\mev)^{-\alpha+1}}
  {(300\kev)^{-\alpha +1} - (50\kev)^{-\alpha+1}}
\equiv C(\alpha)\,,\label{C_alpha}
\eeq
which is valid for all GRBs assumed to have the same integral spectral
index $\alpha$. 

\para 
Based on the above considerations, we shall assume the following LF for
GRBs: 
\beq
\frac{dN}{dL_H}\propto \left\{ \begin{array}{ll}
L_H^{-1.7} & \mbox{for $L_H < L_{H*}$}\\
L_{H*}^{1.1}L_H^{-2.8} & \mbox{for $L_H\geq L_{H*}\,,$}
\end{array}
\right. \label{LF_H}
\eeq
where the break luminosity $L_{H*}$ is given by $L_{H*}=C(\alpha) L_{L*}$,
with $L_{L*}\simeq 2\times10^{52}\erg/\sec$ being the break of the LF
derived in Ref.~\cite{deng}. 

\para
Below, we shall often drop the subscript $H$ in $L_H$, and so, unless
otherwise specified, $L$ will mean $L_H$. 

\para
For the rate-density of GRBs (number of bursts per unit volume and unit
time) in the Universe as a function of redshift, we will assume it to be  
proportional to $(1+z)^{2.5}$ \cite{deng} for the redshift range of our
interest (up to $z\sim 4$). This redshift distribution
follows star formation rate (SFR)~\cite{sfr} for $z < 2$, but unlike the
observed SFR based on optical observations, which generally yield an SFR
that either levels off or falls with redshift for $z>2$, the above GRB
rate increase monotonically with $z$ at least up to $z\sim 5$ \cite{deng}
and possibly even up to $z\sim 10$ \cite{fen}. 
The suggestion~\cite{fen,deng} is that the SFR based on optical
observations may have underestimated the true SFR at high redshifts
because of cosmological reddening effects, and since gamma rays are not
subject to the reddening effect, the GRB rate may indeed reflect the true
SFR at high redshifts provided, of course, the physical connection between
the GRB phenomenon and the star formation process is clearly understood by
future studies. 

\para 
With the LF and rate-density (i.e., redshift distribution) of GRBs
specified above, we now proceed to calculate the expected average rate of
detection of TeV GRBs in the muon detectors. In doing this we follow the 
formalism described in Ref.~\cite{vernetto}. 

\subsection{Calculation of the Rate of TeV GRBs}
The first step is to calculate the redshift up to which a GRB of a given
$L$, $\Egammamax$, $\alpha$ and ${\delta}t$ will be detectable at a zenith 
angle $\theta$ in a given detector. By detectable we mean the statistical
significance of the signal, i.e., the signal to square root of noise
ratio, $S/\sqrt{N}$, is larger than some preassigned value. 
\newpage
\begin{center}
TABLE III
\vskip 0.5cm
\begin{tabular}{|c|c|}\hline
Luminosity ($L$)& Redshift up to which\\
(ergs/sec)& the GRB is detectable\\
\hline
$10^{51}$& 0.057\\
$10^{52}$&0.091\\
$10^{53}$&0.129\\
$10^{54}$&0.168\\
$10^{55}$&0.207\\
$10^{56}$&0.245 \\
\hline
\end{tabular}
\end{center}
\vskip 1cm

\para 
Table III illustrates the values of maximum redshifts up to which GRBs of
various $L$ are detectable by AMANDA at zenith angle $\theta=0^{\circ}$ 
with $S/\sqrt{N}\geq1.5$, assuming $\Egammamax=10\tev$, $\alpha=0.8$ and  
${\delta}t=20 \sec$, with absorption of TeV photons taken into account. 
Fig.~9 shows this in more details for three different values of the
integral spectral index $\alpha$. The regions below
the curves in this figure are the allowed regions for positive detections 
with $S/\sqrt{N}\geq1.5$.

\para 
Next, we calculate the fraction, $f(L,\cos\theta)$, of all bursts of
given luminosity $L$ that are detectable at a zenith angle $\theta$ in the
sense defined above. To do this, we 
first define, for a burst of luminosity $L$ at redshift $z$ at a
zenith angle $\theta$ with respect to a given detector, a quantity
$J(L,z,\cos{\theta})$ such that $J(L,z,\cos{\theta})=1$ 
if the burst is detectable and    
$J(L,z,\cos{\theta})=0$ if it is not. With the redshift distribution of
GRBs as specified above, we can write 
\beq
f(L,\cos{\theta})=\frac{\int_{0}^{z_{\rm max}} J(L,z,\cos{\theta}) 
(1+z)^{2.5}\, \frac{dV}{dz}
dz}{\int_{0}^{z_{\rm max}}(1+z)^{2.5}\, \frac{dV}{dz}dz}\,,\label{f}
\eeq
where $z_{\rm max}$ is the maximum redshift of the GRBs in the Universe, 
and $dV(z)$ is the volume element of the Universe between redshifts $z$
and $z+dz$, which is given by~\cite{charlton} 
\beq
\frac{dV}{dz}=\frac{4\pi}{(1+z)^{3}}
\frac{d}{dz}\left[r^3(z)/3\right]\,.\label{dvdz}
\eeq
In the above expression $r(z)$ is the comoving distance as defined in
Eq.~(\ref{r_z}).

\para 
In our numerical calculations we shall take $z_{\rm max}=4$. 

\para
Finally, folding the fraction $f$ with the LF given in Eq.~(\ref{LF_H}) 
and
the zenith angle distribution of the bursts, $dS(\cos\theta)/d\cos\theta$,
we write the expected TeV GRB rate in muon detector, $R_{\tev}$, as  

\beq
R_{\tev}=R_{\rm total}\int_{L_{\rm min}}^{L_{\rm max}} \int_{0.5}^{1}
\frac{dN(L)}{dL}f(L,\cos{\theta})
\frac{dS(\cos{\theta})}{d\cos{\theta}}
dL d(\cos{\theta})\,,\label{R_tev}
\eeq 
where $dN(L)/dL$ is now the normalized LF and $R_{\rm total}$ is the total
observed GRB rate, and we assume that the detector can view GRBs with good
efficiency between zenith angles $\theta=0^{\circ}$ 
and $\theta=60^{\circ}$. We shall assume a uniform zenith angle
distribution for the bursts and set $dS(\cos\theta)/d\cos\theta = 1/2$. 
Also we take $L_{\rm min}=10^{51}\erg/\sec$ and 
$L_{\rm max}=10^{56}\erg/\sec$. 

\para 
BATSE observes about 1 burst per day with a detection efficiency of 0.3. 
Thus, the GRB rate deduced from BATSE observations is $R_{B}\sim
1000\yr^{-1}$. Assuming that BATSE
can observe GRBs up to a maximum redshift $z_{max}=4$ we set
$R_{\rm total}=R_{B}$.  

\para
We have calculated $R_{\tev}$ for AMANDA for a wide range of values of the
intrinsic parameters specifying the bursts. These rates turn out to
be rather low. For example, with $\Egammamax=10\tev$, $\alpha=0.8$ and
${\delta}t=20 \sec$, and requiring $S/\sqrt{N}\geq1.5$ (the same parameter
set as used in obtaining the numbers in Table III), we get $\sim$ 1
detectable TeV GRB in 25 years in AMANDA. The rate depends somewhat on
$\alpha$; for example, for $\alpha=0.5$ with other parameters kept
same as above, the rate improves to $\sim$ 1 burst in 20 years, while for
$\alpha=1$, the rate is $\sim$ 1 in 40 years. In ICECUBE, assuming its 
effective area will be about 10 times larger, the above expected rates
will be correspondingly larger by the same factor, assuming same threshold
energy for muon detection as in AMANDA. 

\para
Note also that the above rates were calculated for only a $1.5\sigma$
detectability; rates for higher-$\sigma$ detectability will be even lower.   

\subsection{Discussion}
The reasons for the rather low detectability rate of TeV photons from GRBs
in existing muon detectors are clear:  
The absorption of TeV photons in the intergalactic infrared background 
allows
detection of TeV photons from only relatively nearby ($z<0.1$) bursts
as clear from Fig.~9, unless the luminosity of the burst is relatively
large ($\gsim10^{53}\erg/\sec$). However, for a LF falling with $L$, 
and a redshift distribution scaling like the SFR (which increases with
$z$), low-redshift and high-luminosity GRBs are relatively rare. 
Nevertheless, as we have demonstrated in Figs.~6--9, occasional 
nearby GRBs of sufficiently long duration, large luminosity and hard
spectrum are likely to be detectable in existing detectors like AMANDA,
and will certainly be detectable in the next generation detectors such as
ICECUBE. 

\section{Summary and Conclusions}
It is possible that GRBs emit not only sub-MeV photons as detected in
satellite-borne detectors, but also higher energy photons extending to TeV
energies. For a GRB photon spectrum falling with energy, as is usually the
case,
the non-detection of TeV photons in satellite-borne detectors could be due
to their limited size. On the other hand, for sufficiently hard photon
spectrum with integral spectral index $\alpha<1$, although the total
number of photons is dominated by those at the low (keV -- MeV) energy
end, the total energy of the burst would be carried away mostly by the few
high (say, TeV)- energy photons. Thus, the actual total energy emitted in
a burst may be much larger than the total energy estimated 
from the observed fluence of sub-MeV photons in the
satellite detectors, which, if true, would have tremendous implications
for the theories of origin of GRBs. It is, therefore, very important to
study possible ways of detecting possible TeV photons from GRBs. 

\para 
Refs.~\cite{hsy,ah} suggested that TeV photons from GRBs might be
detectable in the existing muon detectors such as AMANDA, Lake Baikal,
and Milagro and future detectors such as the proposed ICECUBE, by
detecting the muons in the atmospheric showers created by TeV photons.
Following the calculations of Ref.~\cite{ah}, we have made detailed
calculations of the detectability of possible TeV photons from
individual GRBs by AMANDA-type detectors, as functions of various 
GRB parameters such as their luminosity, duration, integral spectral
index, and redshift. While our calculations generally yield lower number
of expected muons per burst than that calculated in Ref.~\cite{ah} for the
same set of GRB parameters, we conclude that sufficiently high
luminosity, long duration, and hard spectrum nearby GRBs would still
be detectable in AMANDA-class detectors and will certainly be detectable 
in next generation detectors such as ICECUBE. We have also estimated the
expected rate of TeV GRB events in these detectors using
recent information on the luminosity function of GRBs and their
redshift distribution in the Universe. Our conservative estimates,
including the effect of absorption of TeV photons in the intergalactic
space, show that while the expected average rate of TeV GRB events in
AMANDA is rather low --- about one GRB in 20 years for sufficiently hard
(integral spectral index $\alpha=0.5$) spectrum, the rate in up-coming
bigger detectors such as ICECUBE could be about 10 times larger, i.e.,
about 1 in every couple of years. 

\section{Acknowledgment}
One of us (NG) wishes to thank IIA, Bangalore for hospitality where a
major part of this work was done. PB acknowledges partial support under
the NSF US-India cooperative research grant \# INT-9714627.

\newpage
\begin{figure}
\epsfig{figure=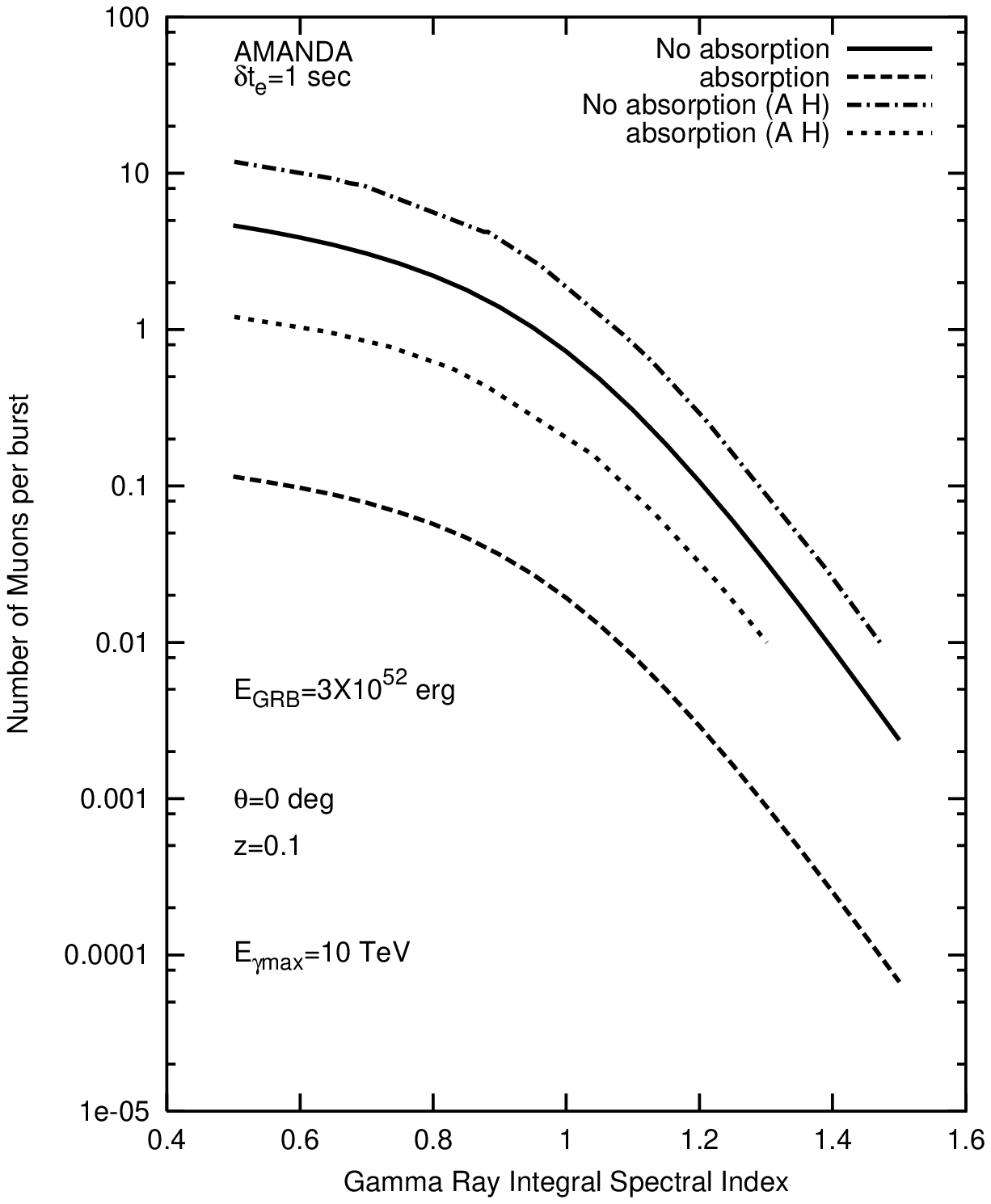,height=9.5cm} 
\caption{Number of muons per burst at zenith angle $\theta=0^\circ$
in AMANDA as a function of the
integral spectral index of the GRB photon spectrum, for a GRB at redshift 
$z=0.1$, total energy $E_{\rm GRB}=3\times10^{52}\erg$, and maximum cutoff
energy $\Egammamax=10\tev$. Curves are shown for with and without
absorption of the photons in the intergalactic infrared field  (IIRF).
Corresponding curves from the calculation of 
Ref.~\cite{ah}, indicated above by A H (their Fig.~1) for the same set
of parameter values as above, are also shown for comparison. In the case
of absorption, our results are almost a factor of 10 below those in  
A H (see text for discussion).}
\end{figure}
\newpage
\begin{figure}
\epsfig{figure=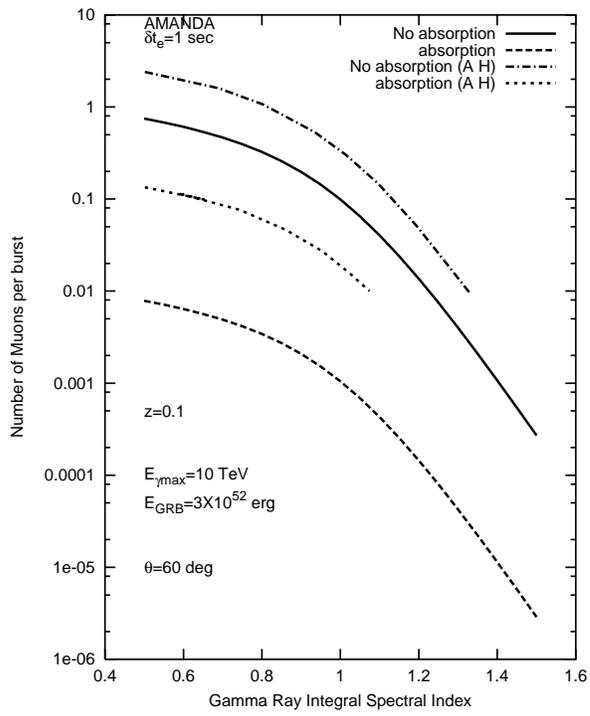,height=9.5cm} 
\caption{Same as Fig.~1, but for zenith angle $\theta=60^\circ$}
\end{figure}
\newpage
\begin{center}
\begin{figure}
\psfig{figure=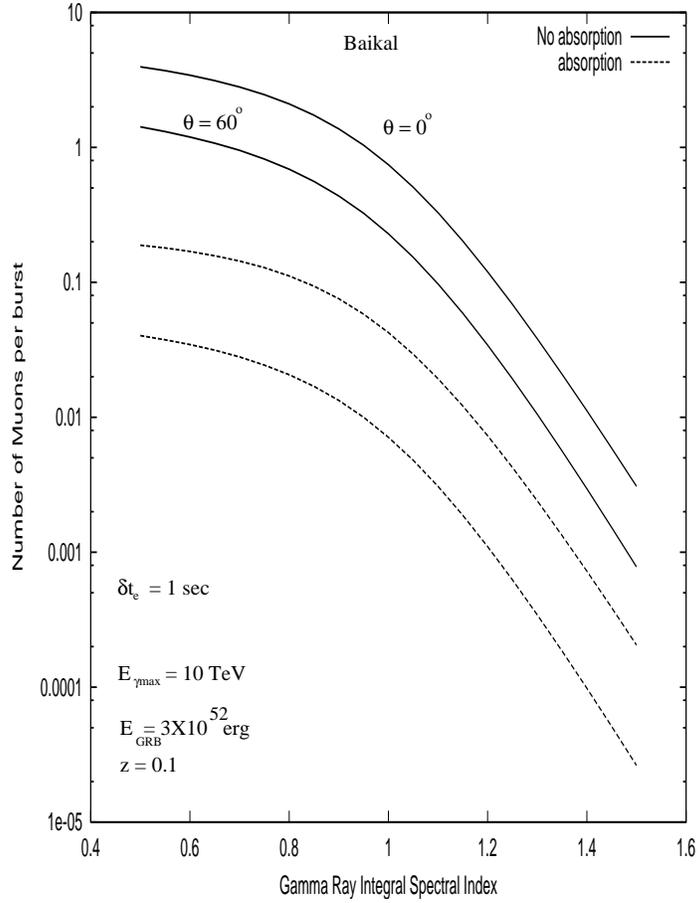,height=9.5cm,angle=270} 
\caption{Number of muons per burst in the Lake Baikal detector as a
function of the integral spectral index of the GRB photon spectrum, for
the same set of GRB parameters as in Fig.~1. Curves are shown for with
and without absorption of the photons in the IIRF and for two different
zenith angles $\theta=0^\circ$ and $\theta=60^\circ$.}
\end{figure}
\end{center}

\begin{center}
\begin{figure}
\psfig{figure=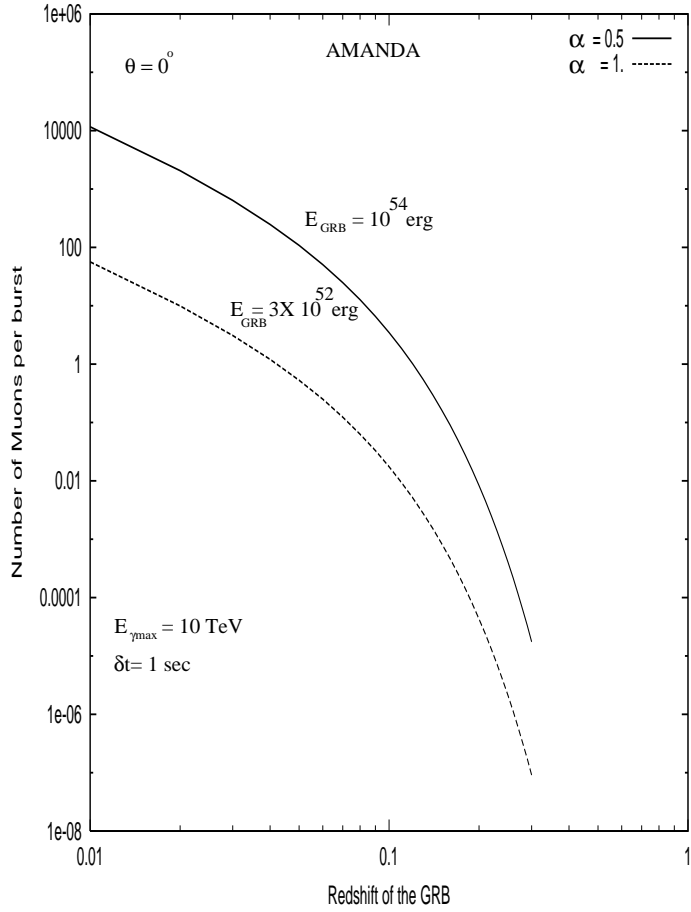,height=9.5cm,angle=270} 
\caption{Number of muons per burst in the AMANDA detector at zenith
angle $\theta=0^\circ$ as a
function of the redshift of the burst for various GRB parameters as
indicated, taking into account the absorption of photons in the IIRF
corresponding to the higher level of IIRF (see text).}   
\end{figure}
\end{center}

\begin{center}
\begin{figure}
\psfig{figure=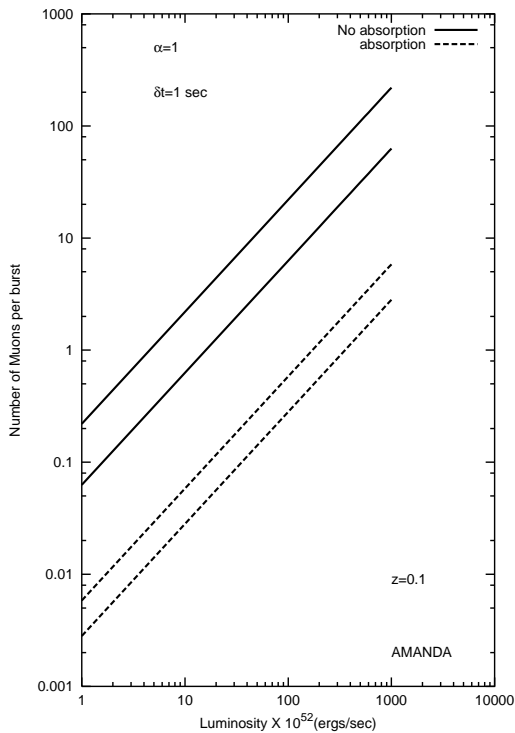,height=9.5cm} 
\caption{Number of muons per burst in the AMANDA detector at zenith
angle $\theta=0^\circ$ as a
function of the luminosity of the burst for various GRB parameters as
indicated. Curves are shown for both with absorption (dashed curves) and
without absorption (solid curves) in the IIRF. The upper solid or
dashed curves are for maximum spectral cutoff of the source at
$\Egammamax=10\tev$ and the lower ones for $\Egammamax=5\tev$.}   
\end{figure}
\end{center}

\begin{center}
\begin{figure}
\psfig{figure=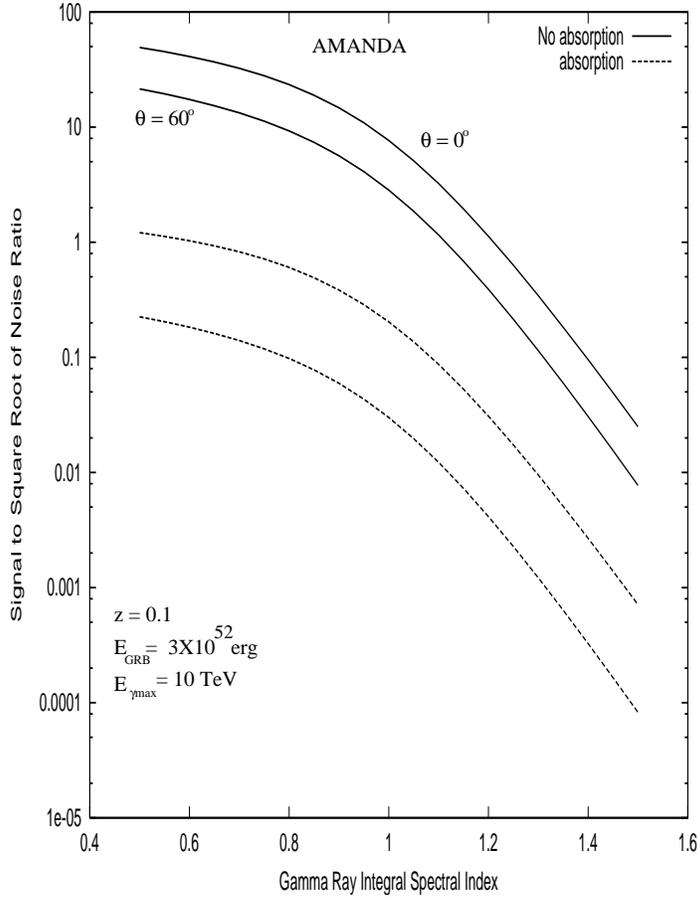,height=9.5cm,angle=270} 
\caption{The signal to square root of noise ratio in the AMANDA detector 
as a function of the integral spectral index of the GRB photon spectrum, 
corresponding to the muon numbers shown in Figs.~1 and 2. } 
\end{figure}
\end{center}

\begin{center}
\begin{figure}
\psfig{figure=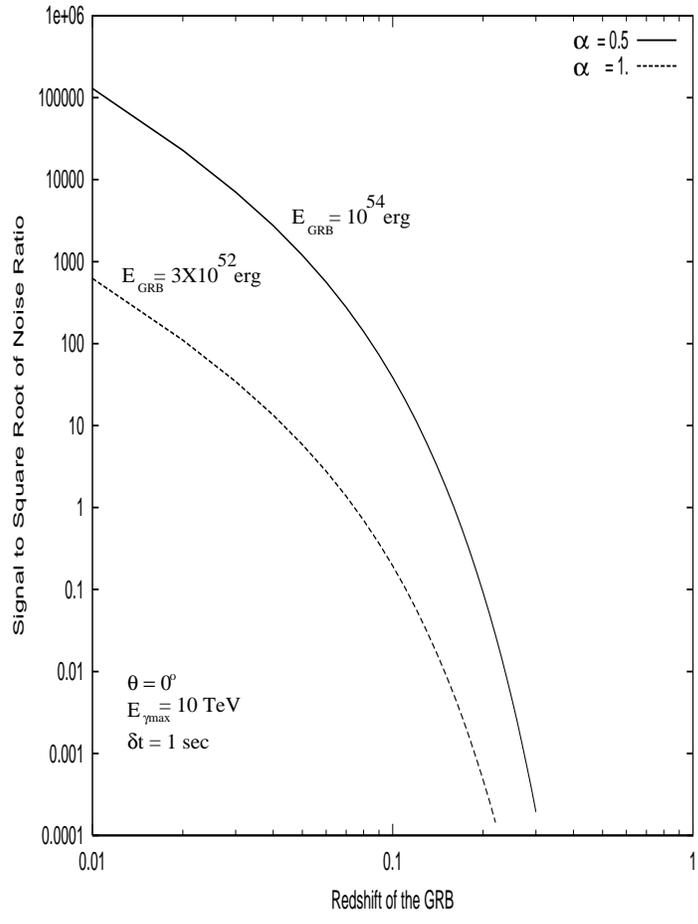,height=9.5cm,angle=270} 
\caption{The signal to square root of noise ratio in the AMANDA detector
as a function of redshift of the GRB, corresponding to the muon numbers
shown in Fig.~4.}
\end{figure}
\end{center}

\begin{center}
\begin{figure}
\psfig{figure=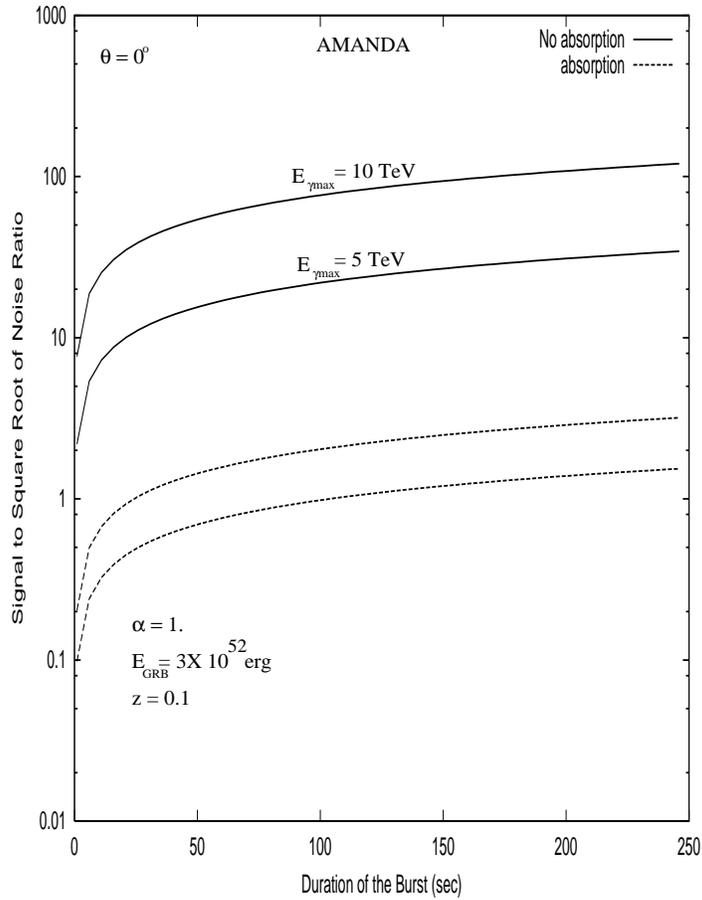,height=9.5cm,angle=270} 
\caption{The signal to square root of noise ratio in the AMANDA detector
as a function of the duration of the GRB.} 
\end{figure}
\end{center}

\begin{center}
\begin{figure}
\psfig{figure=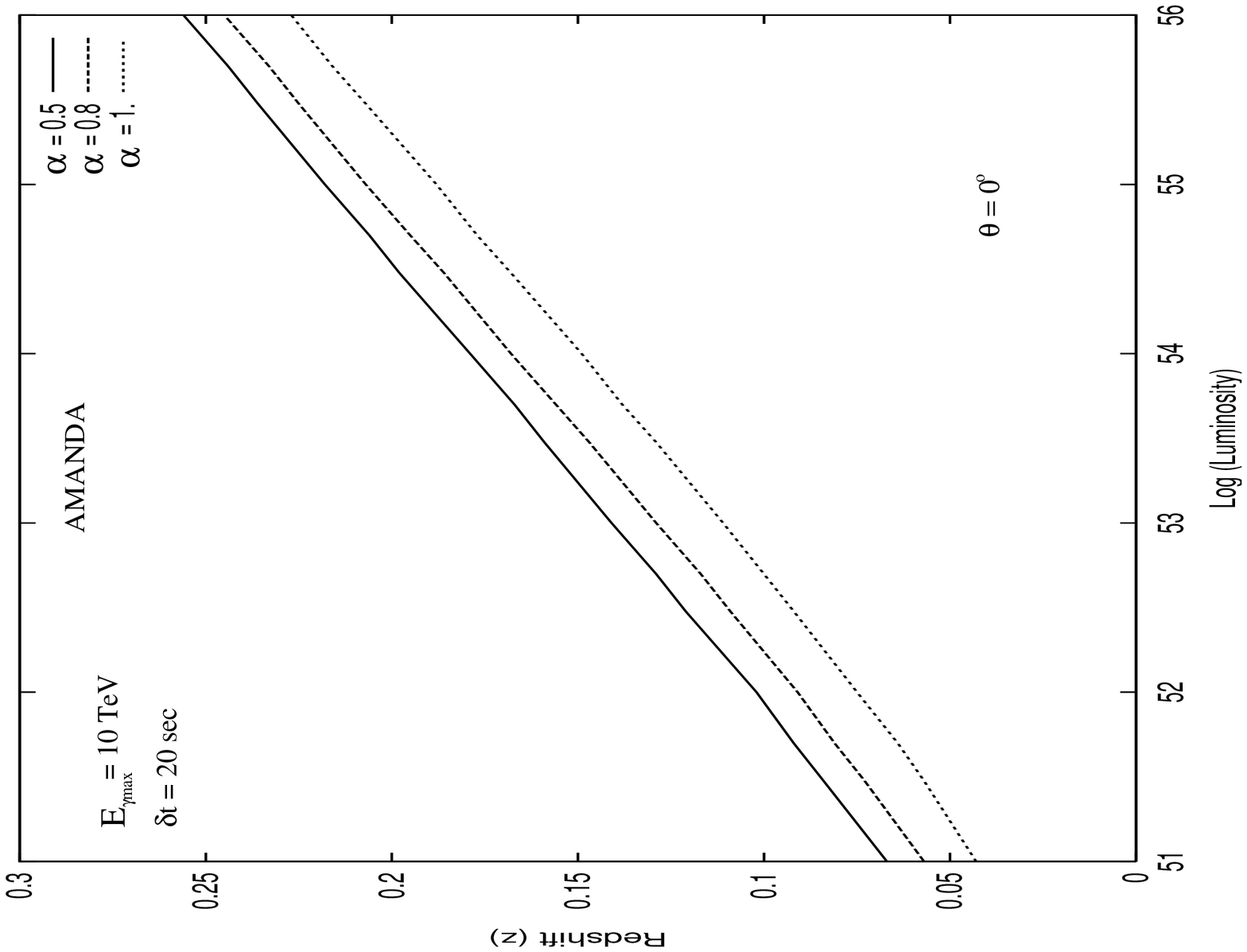,height=9.5cm,angle=270} 
\caption{The maximum redshift of a GRB that can be detected with signal to
square root of noise ratio $S/\sqrt{N}\geq1.5$ in AMANDA as a function of
the luminosity of the burst for several choices of integral spectral
index $\alpha$ and for the indicated values of $\Egammamax$ and burst
duration. The regions below the respective curves are the allowed regions
for positive detection.}  
\end{figure}
\end{center}

\newpage
\begin{thebibliography}{99}
\bibitem{grb_review} T. Piran, Phys.~Rep. {\bf 314}, 575 (1999). 
\bibitem{hurley} K. Hurley et al, Nature {\bf 372}, 652 (1994). 
\bibitem{ong} R.A. Ong, Phys.~Rep. {\bf 305}, 95 (1998). 
\bibitem{tibet} M. Amenomori et al, Astron.~Astrophys. {\bf 311}, 919
(1996). 
\bibitem{hegra} L. Padilla et al, Astron.~Astrophys. {\bf 337}, 43 (1998). 
\bibitem{milagro} R. Atkins et al, Astrophys.~J. {\bf 533}, L119
(2000); also see 
${\sf http://www.lanl.gov/milagro/the_milagro_collaboration.html}$.  
\bibitem{stecker-dejager} F. W. Stecker and O.~C.~de Jager,
Astron. Astrophys. {\bf334}, L85 (1998). 
\bibitem{totani1} T. Totani, Astrophys. J {\bf509}, L81 (1998). 
\bibitem{totani2}T. Totani, Astrophys. J. {\bf 536}, L23 (2000). 
\bibitem{grb-uhecr} E.~Waxman, Phys. Rev. Lett. {\bf 75}, 386 (1995);
M.~Milgrom and V.~Usov, Astrophys.~J. {\bf 449}, L37 (1995); M.~Vietri,
Astrophys.~J. {\bf 453}, 883 (1995). 
\bibitem{vazquez} R. A. Vazquez, astro-ph/9810231. 
\bibitem{totani3} T. Totani, Astropart. Phys. {\bf11}, 451 (1999). 
\bibitem{amanda} F. Halzen (for AMANDA Collaboration)
Nucl.~Phys.~Proc.~Suppl. {\bf 77}, 474 (1999);  E.~Andres et al (AMANDA
collaboration), Astropart.~Phys. {\bf13}, 1 (2000).  
\bibitem{baikal} Ch. Spiering (for BAIKAL Collaboration),
Prog.~Part.~Nucl.~Phys. {\bf 40}, 391 (1998); V.~A.~Balkanov et al (BAIKAL
Collaboration), Astropart.~Phys. {\bf14}, 61 (2000). 
\bibitem{icecube} The ICECUBE proposal: see ${\sf 
http://pheno.physics.wisc.edu/icecube/}$. 
\bibitem{hsy} F. Halzen, T. Stanev, and G. Yodh, Phys.~Rev.~D {\bf 55},
4475 (1997). 
\bibitem{ah} J. Alvarez-Mu\~{n}iz and F. Halzen, Astrophys. J. {\bf521}, 
928 (1999). 
\bibitem{mtw} C.W. Misner, K.S. Thorne and J.A. Wheeler, {\it Gravitation} 
( W.~H.~ Freeman, San Francisco, 1973).  
\bibitem{muon} F. Halzen, K. Hikasa and T. Stanev, Phys.~Rev. D {\bf 34}, 
2061 (1986). 
\bibitem{fen} E. E. Fenimore and E. Ramirez-Ruiz, astro-ph/0004176. 
\bibitem{norris} J.P. Norris, G. Marani and J. Bonnell, Astrophys.~J. {\bf
534}, 248 (2000). 
\bibitem{deng} B. E. Schaefer, M. Deng and D.L.~Band, astro-ph/0101461. 
\bibitem{sfr} C.C. Steidel et al, Astrophys.~J. {\bf 519}, 1 (1999). 
\bibitem{vernetto} S. Vernetto, Astropart. Phys. {\bf 13}, 75 (2000).     
\bibitem{charlton} J.C. Charlton and M.S. Turner, Astrophys.~J. {\bf 313}, 
495 (1987). 
\end{thebibliography}
\end{document}